\documentclass[prl,aps,twocolumn,floatfix,showpacs]{revtex4}
\usepackage{graphicx,graphics,psfrag,amsmath,calc,amssymb}

\begin{document}
\title{Exploring the order parameter symmetry of $p$-wave Fermi condensates}
\author{Wei Zhang}
\affiliation{FOCUS center and MCTP, Department of Physics,
University of Michigan, Ann Arbor, MI 48109}
\author{C. A. R. S\'{a} de Melo}
\affiliation{School of Physics, Georgia Institute of Technology,
Atlanta, GA 30332}

\date{\today}

\begin{abstract}
We discuss the time-of-flight expansion of dilute $p$-wave Fermi
condensates on the BEC side of Feshbach resonances, as a way to extract information
about the order parameter symmetry for superfluidity. We show that the
cloud profile is in general sensitive to
the interaction strength between fermions, the magnitude and
direction of external magnetic fields, and to the angular momentum
projection of the order parameter. In particular, due to the
anisotropic nature of $p$-wave interactions we show that the
time-of-flight expansion of a $p$-wave superfluid is anisotropic
even if the superfluid is confined to a completely isotropic trap,
unlike the case of Bose or $s$-wave Fermi condensates, which under
the same circumstances expand isotropically. Furthermore, we demonstrate
that expanding $p$-wave superfluids released from axially symmetric traps
experience anisotropy inversions, where the aspect ratio between the
axial and radial directions change during expansion, while the axial
symmetry can also be lost reflecting the spatial anisotropy of the
underlying interaction.
\end{abstract}

\maketitle

One of the next frontiers of research in ultra-cold atoms is the
search for superfluidity in higher angular momentum channels, which
are quite rare in nature. With the exception of liquid $^3$He, there
are no known neutral superfluids with angular momentum in the
$p$-wave channel or higher. In addition, there are very few known
examples of $p$-wave and $d$-wave superconductors (charged
superfluids) in standard condensed matter physics, where the degree
of control of the interactions is not existent. In contrast, for
ultra-cold atoms, interactions in higher angular momentum channels
can be tuned via the appropriate choice of Feshbach resonances.

In general, a key point for the understanding of superfluidity is
the determination of the order parameter symmetry, and some
ingenuity is usually necessary to extract this information
experimentally in standard condensed matter systems. However, for
ultra-cold atoms, time-of-flight measurements can provide valuable
information about their superfluidity and the corresponding order
parameter symmetry, as discussed below.

Time-of-flight experiments are powerful methods to study ultra-cold
atoms, and have been applied to confirm the existence of
Bose-Einstein condensation (BEC) and to explore superfluid dynamics
of Bose systems~\cite{anderson-95, davis-95, ohara-02, regal-03,
bourdel-03, schunck-07}. Furthermore, this type of measurements were
used to probe signatures for superfluidity of fermions with $s$-wave
interactions~\cite{ohara-02, regal-03, bourdel-03}.

Time-of-flight techniques were also used to detect $p$-wave Feshbach
resonances of ultra-cold fermions in harmonic
traps~\cite{regal-03b,ticknor-04,zhang-04,schunck-05} and in optical
lattices~\cite{gunter-05}, and paved the way for the exploration of
$p$-wave Fermi superfluidity, where fermions are paired in the
angular momentum channel $\ell =1$. Using these observed resonances,
$p$-wave molecules have been produced in the
laboratory~\cite{jin-07, fuchs-08, inada-08} for fermion isotopes
$^{40}$K and $^{6}$Li, bringing the experimental community a step
closer to realizing $p$-wave Fermi superfluidity of ultra-cold
atoms. In addition, other Fermi atoms like $^{173}$Yb have also been
cooled to quantum degeneracy~\cite{fukuhara-07}, and are likely to
possess $p$-wave Feshbach resonances, thus permitting the study of
$p$-wave molecules or superfluids in non-alkali fermions.

In contrast to their $s$-wave counterparts, $p$-wave Feshbach
resonances have characteristic features including splitting of peaks
depending on hyperfine (pseudospin) states as in $^6$Li and $^{40}$K
(i.e., $\vert 11\rangle$, $\vert 12 \rangle + \vert 21 \rangle$, and
$\vert 22
\rangle$)~\cite{regal-03b,ticknor-04,zhang-04,schunck-05,gunter-05},
and splitting of peaks depending on angular momentum projections as
in $^{40}$K (i.e., the magnetic quantum number $m_\ell =0$ or $\pm
1$)~\cite{ticknor-04,gunter-05}. Hence, these splitting may allow
the separate tuning of $p$-wave scattering parameters in different
pseudospin and/or $m_\ell$ channels, such that $p$-wave interactions
may be anisotropic in both pseudospin and angular momentum states.

Theoretical studies of $p$-wave systems focusing on the evolution
from Bardeen-Cooper-Schrieffer (BCS) to BEC superfluidity have
emphasized the existence of quantum phase
transitions~\cite{botelho-05, gurarie-05}. However, these works
dealt only with thermodynamic properties, while here we study the
time-of-flight expansion of ultra-cold fermions with $p$-wave
interactions and focus on the BEC side of Feshbach resonances, where
the critical temperatures are higher. Our main conclusions are as
follows. We show that the cloud profile reveals the anisotropic
nature of the $p$-wave interaction and provide valuable information
about the order parameter for $p$-wave superfluidity. In particular,
we find that a completely isotropic cloud experiences anisotropic
expansion driven by $p$-wave interactions, in contrast to the
isotropic expansion of Bose and $s$-wave Fermi systems under the
same circumstances. For a cigar-shaped cloud, we find that the
aspect ratio between the axial and radial directions of a $p$-wave
Fermi condensate undergoes an anisotropy inversion driven by the
anisotropy of the harmonic potential, like in Bose and $s$-wave
Fermi systems. However, the axial symmetry of $p$-wave Fermi
condensates may be broken during the expansion reflecting the
spatial anisotropy of the underlying interaction, in sharp contrast
to Bose and $s$-wave Fermi systems, where the axial symmetry is
always preserved.

To describe the time-of-flight expansion of $p$-wave Fermi
condensates, we consider dilute fermions with mass $m$ in a single
hyperfine state (pseudospin). The time-dependent Hamiltonian is
(with $\hbar = k_B = 1$)
\begin{eqnarray}
\label{eqn:hamiltonian}
&&{\cal H} (t) = \int d {\bf r}\bigg\{ \psi^\dagger({\bf r},t)
\left[-\frac{\nabla_{\bf r}^2}{2m} + U_{\rm ext}({\bf r}, t) \right]
\psi({\bf r},t)
\nonumber \\
&&
- \frac{1}{2}\int d{\bf r}' \psi^\dagger({\bf
r},t) \psi^\dagger({\bf r}',t) V({\bf r}-{\bf r}') \psi({\bf r}',t)
\psi({\bf r},t) \bigg\},
\end{eqnarray}
where $\psi^\dagger$ ($\psi$) are creation (annihilation) operators,
$U_{\rm ext}({\bf r}, t)$ is the time dependent trapping potential,
and $V({\bf r}-{\bf r}')$ denotes the fermion-fermion interaction.
The superfluid state in a harmonic trap is characterized by two
length scales: the coherence length $\xi$ and the length scale $R_U$
over which $U_{\rm ext}$ varies. In the limiting case where the
trapping potential varies slowly in comparison to the coherence
length ($R_U \gg \xi$), a semiclassical approximation is applicable
such that one can first consider a dilute Fermi gas in free space
and add the harmonic potential later. With this approximation, the
free space Hamiltonian takes the form
\begin{equation}
{\cal H} = \sum_{\bf k} \xi({\bf k}) \psi_{\bf k}^\dagger \psi_{\bf k}
+ \frac{1}{2}\sum_{{\bf k}, {\bf k}^\prime, {\bf q}} V({\bf k}, {\bf k}^\prime)
B^\dagger_{{\bf k}, {\bf q}} B_{{\bf k}, {\bf q}},
\end{equation}
where $\xi({\bf k}) = k^2/(2m) - \mu$ is the fermion dispersion
shifted by the chemical potential $\mu$, and $ B^\dagger_{{\bf k},
{\bf q}} = \psi_{{\bf k}+{\bf q}/2}^\dagger \psi_{-{\bf k}+{\bf
q}/2}^\dagger $ is the creation operator of a pair with center of
mass momentum ${\bf q}$ and relative momentum $2{\bf k}$. Here, the
momentum space interaction $V({\bf k}, {\bf k}^\prime)$ is the
Fourier transform of $V( {\bf r} -{\bf r}' )$, and can be
approximated by a separable form $V({\bf k}, {\bf k}^\prime) = - 4
\pi \sum_{m_\ell} g_{m_\ell} \Gamma(k) \Gamma(k^\prime)
Y_{1,m_\ell}^*(\hat {\bf k}) Y_{1,m_\ell}(\hat {{\bf k}^\prime})$,
where $\Gamma (k) = k k_0 / (k_0^2 + k^2)$ is the symmetry function
and $Y_{1, m_\ell}$ are spherical harmonics. The momentum scale
$k_0$ is determined by the effective range $R_0 \approx 1/k_0$ of
the interaction and the diluteness condition ($n R_0e^3 \ll 1$)
requires $(k_0 / k_F)^3 \gg 1$, where $n$ is the particle density
and $k_F$ is the Fermi momentum. The interaction can be related to
the scattering volume $a_p$ through the T-matrix~\cite{iskin-06b}.

We introduce the bosonic field operator
$b_{m_\ell}^\dagger=\sum_{\bf k} \Gamma(k) Y_{1,m_\ell}^*(\hat {{\bf
k}}) \psi_{{\bf k}+{\bf q}/2}^\dagger \psi_{-{\bf k}+{\bf
q}/2}^\dagger$, its corresponding auxiliary fiend
$\Delta^*_{m_\ell}$ and integrate out the fermions to derive the
Ginzburg-Landau (GL) effective action
%
$$
S_{\rm GL} = S_0 + \sum_{{\bf q}, \nu,
m_\ell, m_\ell^\prime} {\cal L}^{-1}_{m_\ell, m_{\ell}^\prime} ({\bf
q}, \nu) \Delta_{m_\ell}^* ({\bf q}, \nu)
\Delta_{m_\ell^\prime}({\bf q}, \nu) + S_4,
$$
%
around the trivial saddle point $\Delta_{m_\ell} = 0$ near the
transition temperature $T_c$, where $S_4$ denotes the quartic terms.
The long wavelength behavior of the system can be described by the
expansion of the static part of ${\cal L}^{-1}$ in powers of ${\bf
q}$, leading to
\begin{equation}
\label{eqn:2order}
{\cal L}_{m_\ell, m_\ell^\prime}^{-1}({\bf q}, 0) = a_{m_\ell, m_\ell^\prime}
+ \sum_{i,j} c_{m_\ell, m_\ell^\prime}^{i,j} \frac{q_i q_j}{2m} + \cdots.
\end{equation}
Similarly, the low frequency time evolution of the system can be
described by the frequency dependent part of ${\cal L}^{-1}$ in
powers of $\nu_0$ after analytic continuation $i \nu \to \nu_0 + i
0^+$. For simplicity, we choose to investigate the cases where the
order parameter $\Delta$ is nonzero only for the channel $m_\ell=0$
or $m_\ell = \pm 1$, but not in a mixed state of $m_\ell=0$ and
$m_\ell = \pm 1$. In this case, the tensor
$c_{m_{\ell}}^{i,j}=c_{m_{\ell}}^{i} \delta_{i,j}$ becomes diagonal
but is in general anisotropic.

Returning to space-time coordinates and adding back the trapping
potential, we obtain the equation of motion for a $p$-wave Fermi
condensate
\begin{eqnarray}
\label{eqn:GLequation}
i \frac{\partial}{\partial t} \Delta_{m_\ell}({\bf r},t)
&=&
\Bigg[
- \sum_{i} \frac{c_{m_\ell}^{i}}{d_{m_\ell}} \frac{\nabla_i^2}{2m}
+ \frac{a_{m_\ell}}{d_{m_\ell}}
\nonumber \\
&& \hspace{-2cm}
+ 2 U_{\rm ext} ({\bf r}, t)
+ \frac{b_{m_\ell}}{d_{m_\ell}} \vert \Delta_{m_\ell} ({\bf r},t) \vert^2 \Bigg]
\Delta_{m_\ell}({\bf r},t),
\end{eqnarray}
where the coefficients $a_{m_\ell}, b_{m_\ell}, c_{m_\ell}$, and
$d_{m_\ell}$ can be obtained accordingly~\cite{iskin-06b}. The
trivial expansion of the normal component is included in the
evolution of $S_0$, and is not discussed here. In the BEC limit, the
ratio $c_{m_\ell}^{i}/d_{m_\ell} = 1/2$, such that
Eq.~(\ref{eqn:GLequation}) reduces the conventional Gross-Pitaevskii
(GP) form for a dilute gas of paired fermions with mass $2m$. For
simplicity and definiteness, we consider the case of a static
harmonic trap $U_{\rm ext}({\bf r},t) = \sum_{j=x,y,z} m \omega_j^2
r_j^2/2$ for $t < 0$ which is turned off at $t = 0$. For $t<0$, the
system is described by
\begin{eqnarray}
\label{eqn:staticGP}
\mu_0 \Delta_{m_\ell}({\bf r}) &=&
\Bigg[
- \sum_{i} \frac{c_{m_\ell}^{i}}{d_{m_\ell}} \frac{\nabla_i^2}{2m}
+ \frac{a_{m_\ell}}{d_{m_\ell}}
\nonumber \\
&& \hspace{-1cm}
+ 2 U_{\rm ext} ({\bf R})
+ \frac{b_{m_\ell}}{d_{m_\ell}}
\vert \Delta_{m_\ell}({\bf r}) \vert^2 \Bigg] \Delta_{m_\ell}({\bf r}),
\end{eqnarray}
where $\mu_0$ is the effective chemical potential. For dominant
effective boson interactions the Thomas-Fermi approximation leads to
$\vert \Delta_{m_\ell} ({\bf r}, 0) \vert = [d_{m_\ell}(\mu_0 - 2
U_{\rm ext}({\bf r}))/b_{m_\ell}]^{1/2}$ for $\mu_0 \ge 2 U_{\rm
ext}({\bf r})$, and $\vert \Delta_{m_\ell} ({\bf r}, 0) \vert = 0$
otherwise. When this approximation fails the initial condition for
the time-of-flight expansion can be obtained by solving the static
equation (\ref{eqn:staticGP}) numerically. Furthermore, since the
trapping potential is separable in the spatial variables $r_j$, a
scale transformation $R_j \to r_j \sqrt{d_{m_\ell}/c_{m_\ell}^j}$
and $\Omega_j \to \omega_j \sqrt{c_{m_\ell}^j/d_{m_\ell}}$ can be
introduced to remove the anisotropy in the gradient term of Eq.
(\ref{eqn:GLequation}), leading to an equation with the conventional
GP form.

\begin{figure}
\psfrag{r}{$r_{xz}$}
\psfrag{lambda}{$\tau$}
\psfrag{0.1}{$\varepsilon=0.1$}
\psfrag{0.3}{$\varepsilon=0.3$}
\centerline{\includegraphics[width=7.0cm]{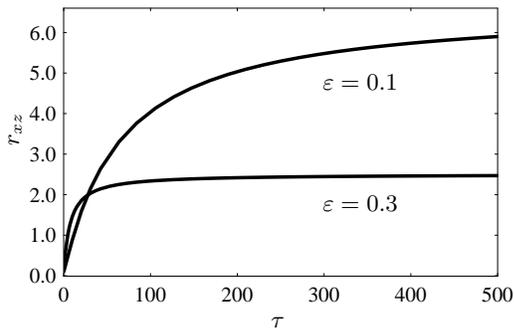}}
\caption{Aspect ratio $r_{xz} = L_x/L_z$ as a function of
dimensionless time $\tau=\omega_\perp t$ during time-of-flight
expansion of a $p$-wave Fermi condensate in the strong attraction
(BEC) limit, for axially symmetric cigar-shaped traps with
$\varepsilon = 0.1$ and $\varepsilon = 0.3$.}
\label{fig:xzratio}
\end{figure}

For $t >0$, we introduce the transformation $R_j(t) = b_j(t)
R_j(0)$, where the scaling factors $b_j(t)$ satisfy $d^2 b_j(t)/dt^2
= \Omega_{j}^2 /[A(t) b_j(t)]$, with $A(t) = b_x(t) b_y(t) b_z(t)$
and initial conditions $b_j(0)= 1$~\cite{castin-96,kagan-96}. Under
the approximation of collisionless hydrodynamics, the time evolution
of the condensate wavefunction
\begin{equation}
\label{eqn:evolution} \Delta_{m_\ell}({\bf R},t) \approx
\frac{\exp[iS({\bf R},t)]}{\sqrt{b_x(t) b_y(t) b_z(t)}}
\Delta_{m_\ell}(\overline{\bf R},0)
\end{equation}
is contained in the scaling factors $b_j (t)$,
scaled coordinates $\overline{R}_k = R_k/b_k(t)$ and phase factor
$S({\bf R}(t), t) = S_0(t) + 2 m \sum_{k} [R_k^2(t)/b_k(t)] d b_k(t)/dt$.
Now, we apply the time evolution results to an axially symmetric
cigar shaped trap with $\omega_x =\omega_y \equiv \omega_\perp \gg
\omega_z$, and consider first the strongly interacting BEC limit. In
the limit where $\varepsilon \equiv \omega_z / \omega_\perp \ll 1$,
the scaling factor $b_j(t)$ can be obtained in powers of
$\varepsilon$, leading to $b_z(t)=1+ \varepsilon^2[\tau
\tan^{-1}\tau - (1/2)\ln(1+\tau^2)] + {\cal O}(\varepsilon^4)$ and
$b_\perp(t) = \sqrt{(1+ \tau^2)} + {\cal O}(\varepsilon^2)$, where
$\tau = \omega_\perp t $ is the dimensionless time. Therefore,
according to Eq. (\ref{eqn:evolution}) the aspect ratio of the cloud
between the radial ($L_x$, $L_y$) and axial ($L_z$) widths is given
by
\begin{equation}
\label{eqn:xzratio}
\frac{L_x(t)}{L_z(t)} = \frac{b_\perp(t) \omega_z}{ b_z(t) \omega_\perp}
= \frac{\varepsilon b_z(t)}{b_\perp (t)}.
\end{equation}
In Fig~\ref{fig:xzratio}, the aspect ratio $r_{xz} \equiv L_x/L_z$
is shown for $\varepsilon = 0.1$ and $\varepsilon = 0.3$, indicating
that the anisotropy of the cloud is inverted during expansion and
that the aspect ratio reaches an asymptotic value significantly
larger than one, similar to the expansion of
Bose~\cite{castin-96,kagan-96} and $s$-wave Fermi~\cite{menotti-02}
condensates.

Notice that in the strong attraction (BEC) limit, fermions form
tightly bound molecules and the internal degrees of freedom of
fermion pairs do not play an important role. Thus, it is not
surprising that the anisotropic nature of the $p$-wave interaction
between fermions does not manifest itself in the behavior of the
condensate. However, as one moves away from the BEC limit towards
unitarity, the average fermion pair size increases, and the internal
structure of pairs can dramatically change the condensate properties
as the pair size becomes comparable to the inter-molecular spacing.
In order to study this regime, we consider a similar case where a
$p$-wave Fermi condensate is initially trapped in an axially
symmetric cigar-shaped potential with $\omega_x = \omega_y \gg
\omega_z$. Furthermore, we impose a magnetic field applied along the
$x$-direction (chosen as the quantization axis) to tune through the
Feshbach resonances. Since resonances in $^{40}$K are
split~\cite{ticknor-04,gunter-05} for different $m_\ell$ states, it
may be possible to adjust the magnetic field such that fermions are
paired in the $m_\ell = 0$ ($p_x$) state only. In this case, the
$p$-wave interaction leads to the formation of $p_x$ symmetry pairs,
which are more strongly correlated along the $x$-direction. As a
consequence, the coefficients $c_{m_\ell=0}^i$ in Eq.
(\ref{eqn:GLequation}) satisfy $ c_{m_\ell=0}^x > c_{m_\ell=0}^y =
c_{m_\ell=0}^z$, hence breaking the axial symmetry in the radial
($x$-$y$) plane. Thus, upon free expansion the cloud looses its
axial symmetry which is initially imposed by the harmonic potential.

\begin{figure}
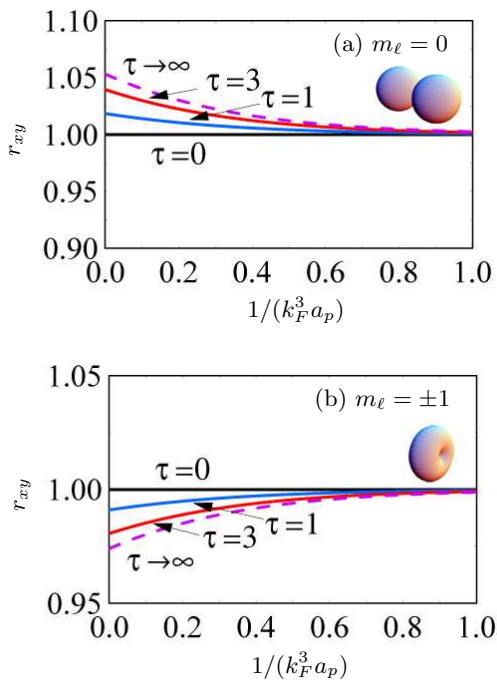

\psfrag{r}{$r_{xy}$}
\psfrag{kfap}{$1/(k_F^3 a_p)$}
\psfrag{a}{(a) $m_\ell=0$}
\psfrag{b}{(b) $m_\ell=\pm 1$}
\centerline{\includegraphics[width=7.0cm]{fig2a-ratio0com2.eps} }
\hspace{1mm}
\centerline{ \includegraphics[width=7.0cm]{fig2b-ratio1com2.eps} }
\caption{Aspect ratio in the radial plane $r_{xy} = L_x/L_y$
as a function of scattering parameter $1/(k_F^3 a_p)>0$
at various time $\tau$ (solid lines) for a $p$-wave
Fermi condensate with (a) $m_\ell =0$ and (b) $m_\ell = \pm 1$
orbital symmetry. Notice that $r_{xy}$ saturates to an asymptotic
value (dashed lines) for $\tau \to \infty$. Insets show the
corresponding pair wave functions.}
\label{fig:xyratio}
\end{figure}

In Fig.~\ref{fig:xyratio}a, we plot the cloud aspect ratio in the
radial plane $r_{xy} = L_x/L_y$ for the $m_\ell = 0$ state as a
function of the $p$-wave scattering parameter $1/(k_F^3 a_p)$ on the
BEC side of the Feshbach resonance. In this plot, we choose the
parameter $k_0/k_F = 5$, which is compatible with a dilute gas of
$^{40}$K with density $n= 10^{14}{\rm cm}^{-3}$. Notice that the
anisotropy disappears in the BEC limit as it must, but becomes more
evident with decreasing interaction strength, leading to a $5\%$
anisotropy near unitarity for sufficiently long times.
Investigations on the BCS side [$1/(k_F^3 a_p)<0$] and at unitarity
require the inclusion of Landau damping which leads to the decay of
Cooper pairs, as well as collisional dynamics of the normal
flow~\cite{gupta-04,jackson-04}, which are beyond the scope of the
present theory. In addition to the anisotropic expansion in the
radial plane which is characteristic of the $p$-wave case, we also
find an anisotropy inversion of the aspect ratio between the axial
and radial directions, which is similar to the cases of Bose and
$s$-wave Fermi condensates.

The anisotropic expansion in the radial plane is better understood
in the language of effective masses. In Eq.~(\ref{eqn:GLequation}),
the coefficient $c_{m_\ell}^i/d_{m_\ell}$ can be absorbed in the
effective mass $M_i = m d_{m_\ell}/c_{m_\ell}^i$. In the BEC limit,
$M_i \to 2m $ is the mass of paired fermions. However, the effective
mass becomes anisotropic for $p$-wave interactions as the system
moves away from the BEC limit, such that for the $p_x$-wave ($m_\ell
= 0$) discussed above, we have $M_x < M_y = M_z$. Thus, it is easier
to accelerate the cloud along the direction of lighter mass $M_x$
such that the cloud expands faster along the $x$-direction than
along the $y$-direction, hence breaking the axial symmetry. In
contrast, if pairing occurs predominantly in the $m_\ell=\pm 1$
channel ($p_y \pm i p_z$-wave), we have $M_x > M_y = M_z$ such that
the cloud expands faster along the $y$ direction, as shown in
Fig.~\ref{fig:xyratio}b. Thus, Figs.~\ref{fig:xyratio}a
and~\ref{fig:xyratio}b help reveal the orbital symmetry of the
$p$-wave order parameter.

A clear distinction between the anisotropy
inversion induced by the harmonic potential and the anisotropic
expansion induced by the $p$-wave interactions is found considering
the time-of-flight expansion for a completely isotropic trap. In
this case, for a $p$-wave Fermi condensate with either $m_\ell = 0$
of $m_\ell = \pm 1$ pairing, the cloud becomes axially symmetric
during expansion and takes a cigar shape (for $m_\ell =0$) or
pancake shape (for $m_\ell = \pm 1$). This is in sharp contrast to
the case of Bose and $s$-wave Fermi condensates, where an initially
isotropic cloud remains alwyas isotropic throughout the expansion.

Before concluding, we would like to stress that although
Eq.~(\ref{eqn:GLequation}) is obtained within the Ginzburg-Landau
scheme for temperatures not too far below $T_c$, our qualitative
results remain valid down to lower temperatures. In particular, the
anisotropy inversion between axial and radial directions is just a
direct consequence of superfluid hydrodynamics, while the
anisotropic expansion in the radial plane is due to anisotropic
$p$-wave interactions. These two qualitative features are
characteristic of $p$-wave Fermi condensates, and are not sensitive
to temperature. Therefore, the qualitative aspects of our results
become more evident at lower temperatures, where a more sizable
condensate fraction emerges (larger bimodal distribution).

In summary, we considered a single component $p$-wave Fermi
condensate and studied the time-of-flight expansion of a cloud
initially trapped in a harmonic potential. We obtained the time
evolution of the condensate on the BEC side of the Feshbach
resonances for temperatures close to the critical temperature.
Starting from a cigar-shaped cloud with axial symmetry, we showed
that the aspect ratio between axial and radial directions presents
an anisotropy inversion during expansion, indicating a
characteristic feature of condensate physics. Furthermore, we found
an anisotropic expansion in the radial plane depending selectively
on the magnetic quantum numbers $m_\ell =0$ and $m_\ell = \pm 1$ of
the $p$-wave interaction. We also emphasized that this anisotropic
expansion is a direct consequence of anisotropic $p$-wave
interactions and occurs most clearly when the harmonic trap is
completely isotropic. In addition, we would like to stress that this
anisotropic expansion should occur not only for $p$-wave, but also
for any higher angular momenta ($d$-wave, $f$-wave, etc...) as
Feshbach resonances are approached from the BEC side. Finally, we
proposed that the orbital symmetry of the order parameter for
$p$-wave condensates can be directly probed through time-of-flight
expansions of harmonically trapped clouds. Potential candidates for
such experiments are $^{40}$K, $^6$Li, and $^{173}$Yb.

We would like to thank NSF for support (Grant N. DMR-0709584).

\end{document}